\documentclass[12pt]{article}
\pdfoutput=1
\usepackage[a4paper]{geometry}
\usepackage{amssymb} 
\usepackage{amsmath}
\usepackage{graphicx}
\usepackage{cite}
\author{\hspace*{-1cm}Thomas R.\ Weikl\\[0.2cm]
 \small Max Planck Institute of Colloids and Interfaces, Department of \\[-0.2cm] 
\small Theory and Bio-Systems, 14424 Potsdam, Germany} 
\date{}
\title{~\\Transition states in protein folding kinetics: Modeling $\Phi$-values of small $\beta$-sheet proteins}
\begin{document}

\maketitle

\begin{abstract}
Small single-domain proteins often exhibit only a single free-energy barrier, or transition state, between the denatured and the native state. The folding kinetics of these proteins is usually explored via mutational analysis. A central question is which structural information on the transition state can be derived from the mutational data. In this article, we model and structurally interpret mutational $\Phi$-values for two small $\beta$-sheet proteins, the PIN and the FBP WW domain. The native structure of these WW domains comprises two $\beta$-hairpins that form a three-stranded $\beta$-sheet. In our model, we assume that the transition state consists of two conformations in which either one of the hairpins is formed. Such a transition state has been recently observed in Molecular Dynamics folding-unfolding simulations of a small designed three-stranded $\beta$-sheet protein. We obtain good agreement with the experimental data (i) by splitting up the mutation-induced free-energy changes into terms for the two hairpins and for the small hydrophobic core of the proteins, and (ii) by fitting a single parameter, the relative degree to which hairpin 1 and 2 are formed in the transition state. The model helps to understand how mutations affect the folding kinetics of WW domains, and captures also negative $\Phi$-values that have been difficult to interpret.
\end{abstract}

\section*{\large Introduction}

How proteins fold into their native 3-dimensional structure remains an intriguing question. Given the vast number of unfolded protein conformations, Cyrus Levinthal argued in 1968 \cite{Levinthal68,Levinthal69} that proteins are guided to their native structure by a sequence of folding intermediates. In the following decades, experimentalists focused on detecting and characterizing metastable intermediates with a variety of methods \cite{Baldwin99a}. While such folding intermediates continue to be of considerable interest \cite{Baldwin99b,Cecconi05}, the view that proteins have to fold in sequential pathways from intermediate to intermediate, now known as `old view' \cite{Baldwin94,Matthews93}, changed in the '90s when statistical-mechanical models demonstrated that fast and efficient folding can also be achieved on funnel energy landscapes that are smoothly biased towards the native state and do not exhibit metastable intermediates \cite{Dill97,Bryngelson95}.  The paradigmatic proteins of this `new view' are two-state proteins, first discovered in 1991 \cite{Jackson91}. Two-state proteins fold from the denatured state to the native state without experimentally detectable intermediate states. Since then, many small-single domain proteins have been shown to fold in two-state kinetics \cite{Jackson98,Fersht99,Grantcharova01}. 

The folding dynamics of two-state proteins is thought to be dominated by a single free-energy barrier, or transition state, between the denatured and native state. This transition state of the protein folding reaction is an instable, short-lived state and cannot be observed directly. Instead, the dynamics of two-state proteins is often explored via mutational analysis \cite{Itzhaki95,Villegas98,Chiti99,Ternstrom99,Kragelund99,Martinez99,Riddle99,Fulton99,Hamill00,Kim00,Mccallister00,Jaeger01,Otzen02,Northey02,Gianni03,Deechongkit04,Garcia04,Anil05,Wilson05,Petrovich06}. In such an analysis, a large number of mostly single-residue mutants of a protein is generated. For each mutant, the effect of the mutation on the folding dynamics is usually quantified by its $\Phi$-value \cite{Matouschek89,Fersht99}
\begin{equation}
\Phi= \frac{R T\ln( k_{\text{wt}}/k_{\text{mut}})}{\Delta G_{N}} \label{phidef}
\end{equation}
Here, $k_{\text{wt}}$ is the folding rate for the wildtype protein, $k_{\text{mut}}$ is the folding rate for the mutant protein, and $\Delta G_{N}$ is the change of the protein stability induced by the mutation. The stability $G_{N}$ of a protein is the free energy difference between the denatured state $D$ and the native state $N$. In classical transition-state theory, 
the folding rate of a two-state protein is proportional to $\exp[-G_T /RT]$, where $G_T$ is the free energy difference from the denatured state to the transition state. It is usually assumed that the prefactor of this proportionality relation does not depend on the mutation. In this notation, $\Phi$-values have the form 
\begin{equation}
\Phi=\frac{\Delta G_{T}}{\Delta G_{N}}  \label{phi2}
\end{equation} 
where $\Delta G_T$ is the mutation-induced change of the free-energy barrier $G_T$. 

The mutational $\Phi$-value data for a protein provide indirect information on its folding dynamics and, therefore, have attracted considerable theoretical interest. The central question is: Which transition transition-state structures and free-energy perturbations are consistent with the experimentally measured $\Phi$-values? In this article, we model $\Phi$-values from detailed mutational analyses \cite{Jaeger01,Deechongkit04,Petrovich06} of two small $\beta$-sheet proteins, the FBP and PIN WW domains. The native structure of these proteins consists of two hairpins forming a three-stranded sheet \cite{Macias00,Ranganathan97} (see Fig.~1). The design principles \cite{Socolich05,Fernandez06} and folding kinetics \cite{Jaeger01,Ferguson02,Nguyen03,Deechongkit04,Deechongkit06,Petrovich06,Jaeger06,Bursulaya99,Ferrara00,Davis02,Karanicolas04,Bruscolini05} of WW domains and other three-stranded $\beta$-sheet proteins have been studied extensively.  Because of their small size and abundance as protein domains, WW domains are important model systems for understanding $\beta$-sheet folding and stability. 
 
Molecular Dynamics (MD) simulations with atomistic models are computationally demanding and in general do not allow direct calculations of folding rates and $\Phi$-values. With additional assumptions on transition states or $\Phi$-values, transition-state conformations have been extracted from MD unfolding trajectories at elevated temperatures \cite{Li96,Day05,Duan05} or constructed from MD simulations that use $\Phi$-values as restraints \cite{Vendruscolo01,Paci02}.\footnote{In statistical-mechanical or Go-type models with simplied energy landscapes, in contrast, folding rates and stabilities for wildtype and mutants and can be easily calculated \cite{Alm99,Munoz99,Galzitskaya99,Clementi00,Guerois00,Alm02,Kameda03}. However, the lack of atomistic detail in these models appears to make it difficult to reproduce detailed mutational data.}
However, for a small, designed three-stranded $\beta$-sheet protein, beta3s, transition-state conformations \cite{Rao05} and $\Phi$-values \cite{Settanni05} have been more rigorously determined from extensive equilibrium folding-unfolding MD simulations. The native structure of beta3s is similar to the structure of WW domains, with two $\beta$-haipins forming an antiparallel three-stranded $\beta$-sheet. Rao et al.\ \cite{Rao05} performed four MD simulations of beta3s at the temperature 330 K with a total length 12.6 $\mu$s,  and observed 72 folding and 73 unfolding events. By identifying clusters of structurally similar conformations that have the probability $p_\text{fold}=0.5$ to fold \cite{Du98,Hummer04,Snow06}, and the same probability to  unfold, Rao et al.\ obtained a transition-state ensemble for beta3s that is ``characterized by the presence of one of the two native hairpins formed while the rest of the peptide is mainly unstructured" \cite{Rao05}. The two $\beta$-hairpins of beta3s thus appear to be cooperative substructures that  are either fully structured or unstructured in the transition state. 

Here, we show that a statistical-mechanical model with a beta3s-like transition-state ensemble in which either hairpin 1 or hairpin 2 are formed leads to an overall consistent interpretation of experimental $\Phi$-values for the FBP and PIN WW domains. In this model, mutations can either affect hairpin 1, hairpin 2, or the small hydrophobic core of the WW domains, which is not yet structured in the transition state. The general form of $\Phi$-values in this model is 
\begin{equation}
\Phi = \frac{\Delta G_T}{\Delta G_N} =\frac{\chi_1\Delta G_1 + \chi_2\Delta G_2}{\Delta G_N} \label{phi}
\end{equation}
where $\chi_1$ is the probability, or fraction, of the transition-state conformation in which hairpin 1 is formed, and  $\chi_2=1-\chi_1$ is the probability of the transition-state conformation with hairpin 2 formed. The mutation-induced changes of the free energy difference between the two transition-state conformations and the denatured state are denoted by $\Delta G_1$ and $\Delta G_2$. The model has just two structural parameters, $\chi_1$ and $\chi_2$, which are obtained from a comparison with the experimental data. Different $\Phi$-values for different mutations simply arise from different `free-energy signatures' $\Delta G_1$, $\Delta G_2$, and $\Delta G_N$ of the mutations. 

In particular, the model reproduces the negative $\Phi$-value for the mutation L36A of the FBP WW domain. 
The mutation destabilizes the native state ($\Delta G_N>0$), but stabilizes hairpin 2 ($\Delta G_2<0$), according to calculations with the empirical force field FOLD-X \cite{Guerois02,Schymkowitz05}. This leads to a negative $\Phi$-value in eq.~(\ref{phi}) since $\Delta G_1$ equals 0 for this mutation. In general, `nonclassical' $\Phi$-values, i.e.~$\Phi$-values that are negative or larger than 1, are obtained in the model if mutations stabilize some structural elements, but destabilize others. The mutation L36A of the FBP WW domain, for example, stabilizes hairpin 2, but destabilizes the hydrophobic core.

Nonclassical $\Phi$-values have been difficult to interpret in the traditional interpretation. In this interpretation, a $\Phi$-value is taken to indicate the degree of structure formation of the mutated residue in the transition-state ensemble T \cite{Fersht99}.  A $\Phi$-value of 1 is interpreted to indicate that the residue has a native-like structure in T, since the mutation shifts the free energy of the transition state T by the same amount as the free energy of the native state N. A $\Phi$-value of 0 is interpreted to indicate that the residue is as unstructured in T as in the denatured state D, since the mutation does not shift the free-energy difference between these two states. $\Phi$-values between 0 and 1 are typically taken to indicate partial native-like structure in T. For a protein with $M$ residues, the traditional interpretation thus implies $M$ structural parameters, the degrees of structure formation of all residues. In contrast, the model presented here has just a single independent parameter, the relative degree to which hairpin 1 and 2 are populated in T. Since degrees of structure formation have to be between 0 (`denatured-like') and 1 (`native-like'), the traditional interpretation can not explain nonclassical $\Phi$-values smaller than 0 or larger than 1. In the present model, nonclassical $\Phi$-values arise from substructural free-changes contributions of different sign (see above). 

We have recently suggested a related, novel model for $\Phi$-values of mutations in $\alpha$-helices of a protein \cite{Merlo05,Weikl07}. The model is based on cooperative helix formation and on splitting mutation-induced free energy changes in helices into secondary and tertiary terms \cite{Weikl07}. The two structural model parameters are the degrees of secondary and tertiary structure formation of the helix in the transition state. For several well-characterized helices \cite{Weikl07}, fitting these two parameters to mutational data leads to a consistent, structural interpretation of the $\Phi$-values. The general conclusion from our helix model and the model for small $\beta$-sheet proteins presented here is that a consistent structural interpretation of $\Phi$-values (i) requires to split up mutation-induced stability changes into free-energy contributions from different substructural elements of a protein, and (ii) can be obtained with few parameters that characterize the degree of structure formation of cooperative elements such as $\alpha$-helices and $\beta$-hairpins in the transition-state ensemble.

\section*{\large Model}

The central assumption of our model is that each of the hairpins is either fully formed or not formed in the transition-state ensemble of the protein. The model has then four states: the denatured state $D$ in which none of the hairpins is formed, a transition-state conformation in which only hairpin 1 is formed,  a transition-state conformation in which only hairpin 2 is formed, and the native state with both hairpins formed. The energy landscape can be characterized by three free-energy differences: The free-energy difference $G_N$ of the native state and the free-energy differences $G_1$ and $G_2$ of the transition-state conformations with respect to the denatured state (see Fig.~2).  

The folding kinetics is described by the master equation 
\begin{equation}
\frac{\text{d} P_n(t)}{\text{d} t} = \sum_{m\neq n} \left[ w_{nm} P_{m}(t) - w_{mn}P_n(t)\right]  \label{master_equation} , 
\end{equation}
which gives the time evolution of the probability $P_n(t)$ that the protein is in state $n$ at 
time $t$. Here, $w_{nm}$ is the transition rate from state $m$ to $n$, defined by   
\begin{equation}
w_{nm}=\frac{1}{t_o} \left(1+e^{G_{n}-G_{m}}\right)^{-1} \label{transrates} 
\end{equation}
provided the states $n$ and $m$ are connected via a single step in which only a single hairpin folds or unfolds \cite{Weikl04}.  For other transitions, i.e. for the direct transition from the denatured state to the native state, and vice versa, the transition rates are zero. Here, $t_o$ is a reference time scale. The transition rates defined above obey detailed balance $w_{nm} P_{m}^e = w_{mn} P_{n}^e$ where $P_{n}^e\sim \exp[-G_n/(R T)]$ is the equilibrium weight for the state $n$. Detailed balance ensures that the system ultimately reaches thermal equilibrium.

The master equation of this four-state model can be solved exactly (see Appendix). For high transition-state barriers $G_1\gg RT$ and $G_2\gg RT$ and a stable native state with $G_N\ll -RT$, the folding rate is given by 
\begin{equation}
k(G_1,G_2) \simeq \frac{1}{2} \left(e^{-G_1/RT} +e^{-G_2/RT}\right) \label{foldingRate}
\end{equation}
in units of $1/t_o$. The folding rate $k$ simply is the sum of the rates for the two possible folding routes on which either hairpin 1 or hairpin 2 forms first. The factor $\frac{1}{2}$ in the equation above arises because a molecule, after reaching one of the barrier states $1$ or $2$, either proceeds to $N$ or returns to $D$, with almost equal probability. 

Mutations correspond to the perturbations of the free energy landscape. A mutation therefore can be characterized by the free energy changes $\Delta G_1$, $\Delta G_2$, and $\Delta G_N$. The folding rate of the mutant then is $k_\text{mut}\equiv k(G_1+\Delta G_1, G_2+\Delta G_2)$. For small perturbations $\Delta G_1$ and $\Delta G_2$, a Taylor expansion of $\ln k_{wt} \equiv \ln k$ to first order leads to
\begin{equation}
\ln k_\text{mut} - \ln k_\text{wt} \simeq  \frac{\partial \ln k}{\partial G_1} \Delta G_1 +  \frac{\partial \ln k}{\partial G_2} \Delta G_2 = -\frac{1}{RT} \left(\chi_1 \Delta G_1 + \chi_2\Delta G_2\right) \label{dlogk}
\end{equation}
with
\begin{equation}
\chi_1 \equiv \frac{e^{-G_1/R T}}{e^{-G_1/R T} + e^{-G_2/RT}} \;\; \text{and} \;\; \chi_2 \equiv \frac{e^{-G_2/R T}}{e^{-G_1/R T} + e^{-G_2/RT}} 
\end{equation}
The two parameters $\chi_1$ and $\chi_2$ quantify the extent to which the transition-state conformation 1 and the transition-state conformation 2 are populated in the transition-state ensemble. From the $\Phi$-value definition (\ref{phidef}) and eq.~(\ref{dlogk}), we obtain the general form of $\Phi$-values given in eq.~(\ref{phi}).

\section*{\large Results}

\section*{\normalsize FBP WW domain}

We first consider the FBP WW domain. Petrovich et al. \cite{Petrovich06} have performed an extensive mutational analysis of the folding kinetics. The $\Phi$-values and stability changes $\Delta G_N$ for the considered mutations are summarized in Table 1, together with an assessment which structural elements are affected by the mutations. This assessment is based on the contact matrix of the FBP WW domain shown in Fig.~3. A black dot at position $(i,j)$ of this matrix indicates that the two amino acids $i$ and $j$ are in contact, i.e.~that the distance between any of their non-hydrogen atoms is smaller than the cutoff distance 4 \AA. Since the contact matrix is symmetric, only one half is represented in Fig.~3. The two contact clusters in the matrix correspond to hairpin 1 and hairpin 2 of the FBP WW domain. The remaining contacts largely correspond to contacts of hydrophobic amino acids, the small `hydrophobic core' of the protein. About half of the mutations performed by Petrovich et al.~affect only either hairpin 1 or hairpin 2. The mutation E7A of amino acid 7, for example, affects the contacts $(7,22)$, $(7,23)$, and $(7,24)$, which are all located in hairpin 1 (see contact map in Fig.~3). The remaining mutations also affect the hydrophobic core, or both hairpins. The mutation Y21A, for example affects the contacts $(8,21)$ and $(9,21)$ in hairpin 1, and the contacts $(21,26)$, $(21,27)$, and $(21,28)$ in hairpin 2. 

To test our model, we first consider all mutations that affect only one of the hairpins. The model predicts that all mutations that affect only hairpin 1 should have the same $\Phi$-value $\chi_1$, and all mutations that affect only hairpin 2 the same $\Phi$-value $\chi_2$. This is a direct consequence of eq.~(\ref{phi}). For mutations that affect only hairpin 1, for example, we have $\Delta G_2=0$ since the mutations don't shift the stability of hairpin 2, and $\Delta G_N=\Delta G_1$ since they also don't affect the hydrophobic core. Eq.~(\ref{phi}) then results in $\Phi=\chi_1$ for these mutations. The $\Phi$-values for the ten mutations that only affect hairpin 1 are plotted in Fig.~4. Except for one clear outlier,\footnote{The data point for the mutation T9A can be confirmed as outlier, e.g., with the Grubb's test \cite{Grubbs69} at the standard significance level of 5 \%. For a set of 10 data points as here, a value of $x$ is an outlier for $z\equiv (x - \bar{x})/\text{SD}> 2.29$ where $\bar{x}$ is the sample mean, and SD the standard deviation. For the mutation T9A with $\Phi$-value $-0.09$, the z-value 2.43 exceeds the critical value 2.29.} 
all $\Phi$-values are centered around the value 0.8, mostly within experimental errors. The mean value of these nine $\Phi$-values (dashed line in Fig.~4) leads to the estimated $\chi_1=0.81\pm 0.06$. The error here is estimated as error of the sample mean. The standard deviation of the $\Phi$-values from the the mean value is $0.18$. The four $\Phi$-values for mutations that affect only hairpin 2 range from 0.08 to 0.39 (see Table 1), with mean value $\chi_2=0.30 \pm 0.08$ and standard deviation 0.16. For both sets of mutations, we thus obtain good agreement with the model. In addition, the sum of the above estimated values for the model parameters $\chi_1$ and $\chi_2$ is close to 1, within the error bounds, which is an additional consistency requirement of the model. The two parameters $\chi_1$ and $\chi_2$ are the fractions to which the two transition-state conformations with either hairpin 1 or hairpin 2 formed are populated. These fractions sum up to 1 since the protein has to take one of the possible routes in the model.  

To include other mutations in the model, we have to estimate the impact of these mutations on the stability of the different structural elements they affect (hairpin 1, hairpin 2, or the hydrophobic core). We use FOLD-X here, a molecular modeling program for the prediction of mutation-induced stability changes \cite{Guerois02,Schymkowitz05}. The FOLD-X force field includes terms for backbone and sidechain entropies, which have been weighted against other terms using experimental data from mutational stability analyses.  FOLD-X has been tested on a set of 1088 point mutants and reproduces the stability changes of 1030 of these mutants with a correlation coefficient of 0.83 and a standard deviation of 0.81 kcal/mol \cite{Guerois02}. With FOLD-X, we calculate the mutation-induced stability changes $\Delta G_N$ for the whole FBP WW domain, and the stability changes $\Delta G_1$ and $\Delta G_2$ of hairpin 1 and 2, depending on whether the mutation affects these hairpins. To calculate $\Delta G_1$ and $\Delta G_2$, we simply `cut out' these hairpins from the PDB structure and estimate the stability of the wildtype and mutant hairpins with FOLD-X (see caption of Table 2 for details). The resulting data are summarized in Table 2. The calculated stability changes $\Delta G_N$ can be directly compared to the experimentally measured stability changes $\Delta G_{N,\text{exp}}$. We include here only mutations in the model for which the FOLD-X predicted stability changes $\Delta G_N$ do not differ by more than a factor 2 from the experimental stability changes $\Delta G_{N,\text{exp}}$. For other mutations, the force-field calculations are unreliable. In Table 2, the calculated stability changes for these mutations are shown in brackets. 

The mutations in Table 2 affect two of the structural elements: The mutations W8F and T13A affect hairpin 1 and the hydrophobic core. For these mutations, we have $\Delta G_2=0$, and $\Phi=\chi_1\Delta G_1/\Delta G_N$ according to Eq.~(\ref{phi}). The mutation Y21A affects both hairpins, hence $\Phi=\left(\chi_1\Delta G_1+\chi_2\Delta G_2\right) /(\Delta G_1+\Delta G_2)$. Finally, the mutations T29G, W30A, and L36V affect hairpin 2 and the hydrophobic core. Therefore, we have $\Delta G_1=0$ for these mutations, and $\Phi=\chi_2\Delta G_2/\Delta G_N$. 

Let us now consider the set of 20 mutations that consists of these 6 mutations that affect two structural elements and the 14 mutations that affect either only hairpin 1 or only hairpin 2. Our model has two parameters, $\chi_1$ and $\chi_2$. However, since $\chi_1 + \chi_2=1$, there is only one independent parameter. We determine this parameter from a least-square fit between the theoretical $\Phi$-value formula given in eq.~\ref{phi} and the experimental $\Phi$-values and obtain the values $\chi_1= 0.77 \pm 0.05$ and $\chi_2=0.23 \pm 0.05$, see Fig.~5. 

\section*{\normalsize PIN WW domain}

Mutational analyses of the PIN WW domain's folding kinetics have been performed by J\"ager et al.\ \cite{Jaeger01} and Deechongkit et al.\ \cite{Deechongkit04}. While J\"ager et al.~have considered standard single-site amino-acid replacements, Deechongkit et al.\  synthesized amid-to-ester mutants that specifically perturb backbone H-bonds.  The experimental $\Phi$-values and stability changes $\Delta G_{N,\text{exp}}$ for these mutations are summarized in Table 3. The synthetic amino acids in the mutations of Deechongkit et al.\ are denoted by lowercase greek letters (last six lines in Table 3). Since these mutations perturb the backbone H-bonds, they only affect either hairpin 1 or hairpin 2, which is indicated in the last column in Table 3. For the mutations considered by J\"ager et al., the affected structural elements are again assessed based on the contact map shown in Fig.~3. We consider here only mutations with stability changes $\Delta G_{N,\text{exp}}>0.8$ kcal/mol. $\Phi$-values of mutations that cause significantly smaller stability changes are often considered as unreliable \cite{Garcia04,Fersht04,DelosRios06} (see also Discussion). 

Seven mutations in Table 3 affect only hairpin 1 of the PIN WW domain. The mean value of the $\Phi$-values for these mutations leads to the estimate $\chi_1=0.69\pm 0.05$. The standard deviation of the $\Phi$-values from the mean is 0.12, which is comparable to the experimental errors. The four $\Phi$-values of the mutations that affect only hairpin 2 have the mean value $\chi_2=0.36\pm 0.05$ and the standard deviation 0.10. In agreement with our model, these estimates for $\chi_1$ and $\chi_2$ again add up to 1, within the statistical errors. In an alternative approach, the values of $\chi_1$ and $\chi_2$ can be obtained from a least-square fit between theoretical and experimental $\Phi$-values (see Fig.~6). From the fit, we obtain $\chi_1=0.67\pm 0.05$ and $\chi_2=1-\chi_1 = 0.33\pm 0.05$, and a Pearson correlation coefficient of 0.85 between theoretical and experimental $\Phi$-values. 

We do not include mutations that affect more than one structural element here since the stability changes estimated with FOLD-X appear to be unreliable. For four of the five mutants, the calculated stability changes $\Delta G_N$ differ by significantly more a factor 2 from experimental values $\Delta G_{N,\text{exp}}$ (data not shown). The stabilities for the PIN WW domain mutants may be more difficult to calculate since they involve a larger range of amino acids, compared to the FBP WW mutants that mostly involve changes to the small amino acids Alanine or Glycine, which can be modeled via simple truncation of sidechains prior to the FOLD-X calculations.

\section*{\large Discussion and Conclusions}

We have modeled $\Phi$-values from extensive mutational analyses of two WW domains based on the central assumption that the transition state ensemble of these proteins consists of two substates in which either hairpin 1 or hairpin 2 are formed. The structural information obtained from the mutational data by fitting a single model parameter is that the transition state ensemble of the FBP WW domains consists to roughly $\frac{3}{4}$ of substate 1 with hairpin 1 formed, and to $\frac{1}{4}$ of substate 2 with hairpin 2 formed. The transitions state ensemble of the PIN WW domain consists to roughly $\frac{2}{3}$ of substate 1, and to $\frac{1}{3}$ of substate 2, according to the model.

In the model, the magnitude of a $\Phi$-value depends on which structural elements are affected, and on the mutation-induced free energy changes of these elements. The mutation E7A of the FBP WW domain, for example, has a relatively large $\Phi$-value since this mutation only affects hairpin 1, which is structured in the dominant substate 1 of the transition state ensemble, whereas the mutation W8F has a relatively small $\Phi$-value since the mutation mainly affects the free energy of the small hydrophobic core, which is not yet formed in the transition state. The model also reproduces the negative $\Phi$-value of the mutation L36A, which results from different signs of the mutations-induced free energy changes $\Delta G_1$ and $\Delta G_N$ in Table 2. According to the free-energy calculations with FOLD-X, the mutation stabilizes hairpin 1 ($\Delta G_1<0$), but has an overall destabilizing effect ($\Delta G_N>0$) since it destabilizes the hydrophobic core.

The deviations between experimental and theoretical $\Phi$-values are within reasonable errors. It has been recently suggested that experimental errors for $\Phi$-values may be underestimated since it is usually assumed that the errors in the measured free energy changes of the transition state and the folded state are independent, which is not the case \cite{Ruczinski06}. In case of the PIN WW domain, we have only considered mutations with stability changes $\Delta G_N > 0.8$ kcal/mol. For mutations that induce significantly smaller stability changes, experimental errors in $\Delta G_N$  may lead to large errors in $\Phi$-values since $\Delta G_N$ constitutes the denominator of the $\Phi$-value defined in eq.~(\ref{phidef}). 

However, the large $\Phi$-values up to 1.8 for three mutations with small stability changes in the loop of hairpin 1 of the PIN WW domain \cite{Jaeger01}, which have not been considered here, may also result from structural rearrangements. J\"ager et al.\ \cite{Jaeger01} have suggested a five-state model with two consecutive transition states. In the first transition state, only the loop of hairpin 1 is formed. Nonclassical $\Phi$-values greater than 1 are obtained in this model for mutations  
that are assumed to shift the free energy of the loop by a larger amount than the free energy of the native state. With the same assumption, large nonclassical $\Phi$-values  in the loop of hairpin 1 are also obtained in the four-state model presented here. For $\chi_1 = 0.67$, for example, a $\Phi$-value of 1.8 is obtained for a mutation in this loop with $\Delta G_1=2.7 \Delta G_N$, according to eq.~(\ref{phi}), since hairpin 2 and, thus, $\Delta G_2$ are not affected by this mutation. Such a situation may result from a structural rearrangement between the transition-state conformation with hairpin 1 formed and the native state.  The structural rearrangement may affect the sidechains in the loop, but should not affect the backbone hydrogen bonds since the $\Phi$-values for the amide-to-ester mutations S16$\sigma$, R17$\rho$, and S19$\sigma$ in this loop are between 0.70 and 0.83 (see Table 3). Within the experimental and statistical errors, these $\Phi$-values are close to $\chi_1 = 0.67$, which is the expected $\Phi$-value for mutations with  $\Delta G_1= \Delta G_N$.

\begin{appendix}
\section*{\large Appendix: Exact solution of the master equation}

The master equation (\ref{master_equation}) can be written in the matrix form: 
\begin{equation}
\frac{\text{d}{\boldsymbol P}(t)}{\text{d} t} = -{\boldsymbol W} \boldsymbol{P}(t) 
\end{equation}
The elements of the vector $\boldsymbol{P}(t)$  are the probabilities $P_n(t)$ that the protein is in state $n$ at time $t$, and the matrix elements of $\boldsymbol{W}$ are given by 
\begin{equation}
W_{nm} = -w_{nm} \hspace{0.3cm}\text{for}\hspace{0.3cm} n\neq m; \hspace{0.5cm} 
W_{nn} = \sum_{m\neq n} w_{mn} . 
\end{equation}
For the model with four states considered here, the matrix $\boldsymbol{W}$ is given by 
\begin{equation}
\boldsymbol{W} \!=\! \frac{1}{t_o} \!
\begin{pmatrix} 
\frac{1}{1+e^{g_1}} {\scriptstyle +} \frac{1}{1+e^{g_2}} & - \frac{1}{1+e^{-g_1}} & - \frac{1}{1+e^{-g_2}} & 0 \\ 
 - \frac{1}{1+e^{g_1}} & \frac{1}{1+e^{-g_1}} {\scriptstyle +}  \frac{1}{1+e^{g_N - g_1}} & 0 &  - \frac{1}{1+e^{g_1 - g_N}} \\
 - \frac{1}{1+e^{g_2}} &  0 & \frac{1}{1+e^{-g_2}} {\scriptstyle +}  \frac{1}{1+e^{g_N - g_2}} &  - \frac{1}{1+e^{g_2 - g_N}} \\
 0 & -\frac{1}{1+e^{g_N - g_1}} & - \frac{1}{1+e^{g_N - g_2}} & \frac{1}{1+e^{g_1 - g_N}} {\scriptstyle +} \frac{1}{1+e^{g_2 - g_N}}
\end{pmatrix}\nonumber
\end{equation}
To simplify the notation, we have used here dimensionless free-energy differences $g_i\equiv G_i/RT$ ($i=1$, 2, or $N$) of the partially folded states 1 and 2 and the native state $N$ with respect to the denatured state. 

The general solution $\boldsymbol{P}(t)$ of the master equation can be expressed in terms of the eigenvalues $\lambda$ and eigenvectors $\boldsymbol{Y}_\lambda$ of the matrix $\boldsymbol{W}$: 
\begin{equation}
\boldsymbol{P}(t)=\sum_{\lambda} c_{\lambda} \boldsymbol{Y}_{\lambda} \exp[-\lambda t] \label{gensol} 
\end{equation}
The prefactors $c_{\lambda}$ in this general solution depend on the initial conditions at time $t=0$. For the $4\times 4$ matrix above, the 4 eigenvalues are given by $\lambda = 0$, $1-q$, $1+q$, and 2, in units of $1/t_o$, with
\begin{equation}
q\equiv \frac{1-e^{g_{N}-g_1-g_2}}{\sqrt{(1+e^{-g_1})(1+e^{-g_2})
   (1+e^{g_{N}-g_1})(1+e^{g_{N}-g_2})}} \label{q}
\end{equation}
Since we have $-1< q< 1$, the three nonzero eigenvalues are positive and describe the relaxation to the equilibrium state of the model (see eq.~(\ref{gensol})). The equilibrium state simply is  $c_o \boldsymbol{Y}_o$ where  $\boldsymbol{Y}_o$ is the eigenvector with eigenvalue 0.

This model exhibits two-state folding kinetics under two conditions.  First, the native state has to be stable, i.e.~the free energy $g_N$ of the native state must be significantly smaller than the free energies of the other three states.  Second, the free energy differences $g_1$ and $g_2$ between the intermediate states and the denatured have be to significantly larger than $RT$. The partially folded states then constitute the transition-state ensemble. Under these two conditions, the three Boltzmann weights $e^{g_{N}-g_1-g_2}$, $e^{g_{N}-g_1}$, and $e^{g_{N}-g_2}$ in eq.~(\ref{q}) are much smaller than 1, and also much smaller than $e^{-g_1}$ and $e^{-g_2}$, which leads to
\begin{equation}
q\simeq \frac{1}{\sqrt{(1+e^{-g_1})(1+e^{-g_2})}}
\end{equation}
For large barrier energies $g_1$ and $g_2$, we have $e^{-g_1}\ll 1$ and $e^{-g_2}\ll 1$, and therefore $(1+e^{-g_1})(1+e^{-g_2})\simeq (1+e^{-g_1} +e^{-g_2})$. If we now use the expansion $(1+x)^{-1/2}\simeq 1- x/2$ with $x =e^{-g_1} +e^{-g_2} \ll 1$, the smallest nonzero relaxation rate, or folding rate, $k\equiv 1-q$ is given by eq.~(\ref{foldingRate}), i.e.~by
$k \simeq \frac{1}{2} \left(e^{-g_1} +e^{-g_2}\right)$ in the notation used in this appendix. The folding rate $k$ is much smaller than the other two relaxation rates $1+q$ and 2, which corresponds to an initial `burst phase'.  \end{appendix}

\clearpage

\begin{table}

\vspace*{-1cm}
\thispagestyle{empty}

\hspace*{2cm}Table 1:  Mutational data for the FBP WW domain 

\begin{center}
\begin{tabular}{cccl}
& & & affected\\[-0.2cm]
mutation & $\Phi_\text{exp}$ & $\Delta G_{N,\text{exp}}$  & elements \\
 \hline
E7A & $0.67\pm 0.21$ & $0.52 \pm 0.16$   & hp 1 \\
W8F & $0.24 \pm 0.03$ & $1.65 \pm 0.16$ & hp 1, hc \\
T9A & $-0.09 \pm 0.04$ & $0.93 \pm 0.09$ & hp 1 \\
T9G &  $0.94 \pm 0.20$ & $0.50 \pm 0.10$   & hp 1 \\
Y11A & $0.55 \pm 0.10$ & $0.63\pm 0.11$ & hp 1 \\
T13A & $-0.03 \pm 0.07$ & $0.81\pm 0.17$ & hp1, hc     \\
T13G & $-0.32 \pm 0.25$ & $0.58\pm 0.22$ & hp 1, hc  \\
A14G & $ 0.69 \pm 0.28$ & $0.50 \pm 0.22$ & hp 1\\
D15A & $0.82 \pm 0.16$ & $0.42 \pm 0.09$ & hp 1 \\
D15G & $0.77 \pm 0.17$ & $0.39 \pm 0.09$ & hp 1 \\
G16A & $1.17 \pm 0.22$ & $1.33\pm 0.27$ & hp 1 \\
T18A & $0.93 \pm 0.27$ & $0.54 \pm 0.17$ & hp 1 \\
T18G & $ 0.73 \pm 0.05$ & $1.14\pm 0.09$ & hp 1 \\
Y19A & $0.11\pm 0.05$  & $0.67 \pm 0.13$ & hp 1, hp 2  \\ 
Y20F & $0.05 \pm 0.16$ & $0.68\pm 0.18$ & hp 1, hp 2, hc    \\
Y21A & $0.28 \pm 0.02$ & $1.70 \pm 0.10$ & hp 1, hp 2  \\
R24A & $0.29 \pm 0.09$ & $0.78 \pm 0.17$ & hp 1, hp 2  \\
T25A & $0.39 \pm 0.04$ & $2.51 \pm 0.18$ & hp 2 \\
T25S & $0.27 \pm 0.03$ & $1.08 \pm 0.09$  & hp 2 \\
L26A & $0.08 \pm 0.08$ & $0.56 \pm 0.12$ & hp 2 \\
L26G & $0.45 \pm 0.04$ & $-1.29 \pm 0.10$ & hp 2 \\
E27A & $0.12 \pm 0.04$ & $1.02 \pm 0.13$ & hp 2, hc  \\
T29G & $0.09 \pm 0.02$ & $1.89 \pm 0.11$ & hp 2, hc  \\
W30A & $0.19 \pm 0.06$ & $0.76 \pm 0.14$ & hp 2, hc \\
L36A & $-0.30 \pm 0.16$ & $0.91 \pm 0.14$ & hp 2, hc   \\
L36V & $-0.13 \pm 0.09$ & $0.53 \pm 0.14$ & hp 2, hc  
\end{tabular}

\end{center}
Experimental $\Phi$-values and stability changes $\Delta G_{N,\text{exp}}$ are from Petrovich et al.\ \cite{Petrovich06}. The information on the structural elements affected by the mutations is derived from the contact map shown in Fig.~3. These structural elements are the hairpin 1 (hp 1), hairpin 2 (hp 2), and the small hydrophobic core (hc) of the protein (see text).
\end{table}

\begin{table}

Table 2:  Experimental and calculated stability changes for mutations of the FBP WW domain that affect several structural elements 

\begin{center}
\begin{tabular}{ccclcc}
mutation & $\Delta G_{N,\text{exp}}$  & $\Delta G_N$ & $\Delta G_1$ & $\Delta G_2$ \\ 
\hline
W8F &  $1.65 \pm 0.16$ & $2.39$ & $0.21$ & -- \\
T13A & $0.81\pm 0.17$ &  $0.69$ &  $0.22$ & -- \\
T13G & $0.58\pm 0.22$ &  $(1.28)$ & $(0.56)$ & --  \\
Y19A &  $0.67 \pm 0.13$ &  $(2.65)$ & $(1.60)$ & $(1.01)$ \\ 
Y20F &  $0.68 \pm 0.18$ &  ($- 0.76$) & ($0.31$) & ($-0.45$) \\
Y21A &  $1.70 \pm 0.10$  & $2.58$ & $0.56$ & $1.42$ \\
R24A &  $0.78 \pm 0.17$ & ($-0.23$)  & ($-0.31$) & ($-0.38$)  \\
E27A &  $1.02 \pm 0.13$ & $(0.17)$ & -- & $(0.17)$ \\
T29G &  $1.89 \pm 0.11$  & $1.47$ & -- & $1.14$ \\
W30A & $0.76 \pm 0.14$ & $1.32$ & -- & $0.53$ \\
L36A &   $0.91 \pm 0.14$  & $0.47$ & -- & $-0.30$ \\
L36V &   $0.53 \pm 0.14$ &  $(0.23)$ & -- & ($-0.34$) \\
\end{tabular}
\vspace*{0.5cm}
\end{center}
Experimental data for the stability changes $\Delta G_{N,\text{exp}}$ are from Petrovich et al.~\cite{Petrovich06}. The stability changes $\Delta G_N$,  $\Delta G_1$, and $\Delta G_2$ for the whole protein and hairpin 1 or 2, respectively, have been calculated with the program FOLD-X \cite{Guerois02,Schymkowitz05}. For mutations to alanine (A) or glycine (G) and the muation W8F, native structures for the mutant proteins have been generated by truncation of atoms. For the mutations Y20F and L36V, mutant structures were generated with the program WHAT IF \cite{Vriend90}. The wildtype structure used in the calculations is model 1 of the PDB structure 1E0L \cite{Macias00}. To calculate $\Delta G_1$ and $\Delta G_2$, substructures consisting of the residues 1 to 24 and 15 to 37 of the PDB structure have been used. The FOLD-X calculations have been performed at the ionic strength 150 mM and temperature 283 K of the experiments \cite{Petrovich06}. Numbers in brackets indicate that the calculated stability changes are not reliable since $\Delta G_N$ differs by more than a factor 2 from $\Delta G_{N,\text{exp}}$.
\end{table}

\clearpage

\begin{table}
\hspace*{2cm}Table 3:  Mutational data for the PIN WW domain 

\begin{center}
\begin{tabular}{ccccl}
&&& affected \\[-0.2cm]
mutation & $\Phi_\text{exp}$ & $\Delta G_{N,\text{exp}}$  & elements\\ 
 \hline
L7A & $0.18\pm 0.07$ & 2.06 &  hc \\ 
R14F & $0.68 \pm 0.11$ & 1.29 & hp 1 \\
M15A & $0.63 \pm 0.14$ & 0.90  &  hp 1 \\  
Y23L & $0.64 \pm 0.08$ & 1.51 &  hp 1, hp 2 \\  
Y24F & $0.52\pm 0.14$ & 0.87 &  hp 1, hp 2 \\  
F25L & $0.49 \pm 0.08$ & 1.69  &   hp 1, hp 2 \\ 
N26D & $0.33 \pm 0.05$ & 2.13 &  hp 1, hp 2 \\ 
T29D & $0.30 \pm 0.07$ & 1.77 &  hp 2 \\ 
A31G & $0.44 \pm 0.06$ & 1.88 &  hp 2, hc \\ 
W34A & $0.36 \pm 0.13$ & 1.12 &  hp 2 \\  
K13$\kappa$ & $0.50 \pm 0.05$ & 1.00 & hp 1\\
S16$\sigma$ & $ 0.70 \pm 0.05$ & 1.39 & hp 1\\
R17$\rho$   & $0.78 \pm 0.11$   & 0.74 & hp 1 \\
S19$\sigma$ & $0.83 \pm 0.04$  & 2.03 & hp 1\\
H27$\eta$ & $0.28 \pm 0.03$ & 1.77 & hp 2\\
S32$\sigma$ & $0.51 \pm 0.03$ & 1.77 & hp 2\\
\end{tabular}
\end{center}
Experimental $\Phi$-values and stability changes $\Delta G_{N,\text{exp}}$ for the mutations L7A to W34A are from J\"ager et al. \cite{Jaeger01}, and for the amid-to-ester mutants K13$\kappa$ to S32$\sigma$ from Deechongkit et al.\ \cite{Deechongkit04}. Here, only mutations with stability change  $\Delta G_{N,\text{exp}}>0.8$  kcal/mol are considered. The structural elements affected by the mutations are assessed from the contact map shown in Fig.~3. These structural elements are the hairpin 1 (hp 1), hairpin 2 (hp 2), and the hydrophobic core (hc) of the protein (see text).
\end{table}

\clearpage

\begin{figure}
\begin{center}
\resizebox{0.9\linewidth}{!}{\includegraphics{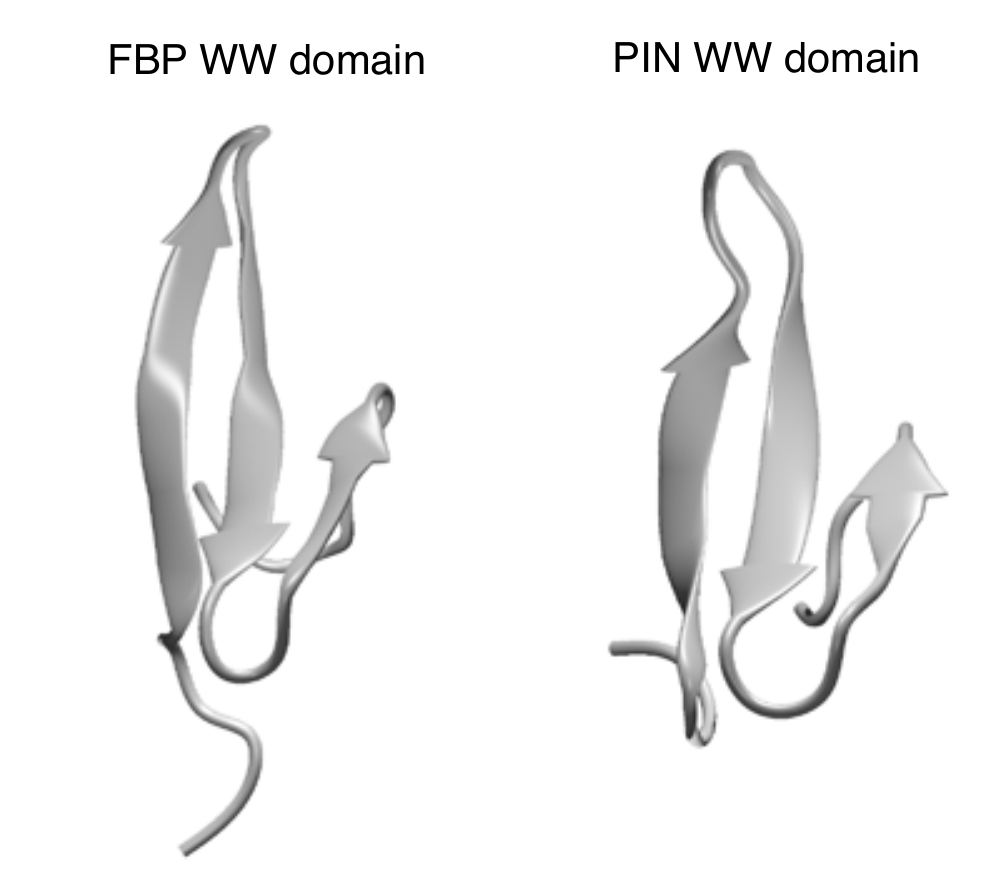}}
\end{center}
\vspace{0.5cm}
\caption{Native structures of the FBP \cite{Macias00} and the PIN WW domain \cite{Ranganathan97}.
The structural representations have been generated with the programs VMD \cite{Humphrey96} and Raster3D \cite{Merritt97}.}
\label{figure_structures}
\end{figure}

\clearpage

\begin{figure}
\begin{center}
\resizebox{0.6\linewidth}{!}{\includegraphics{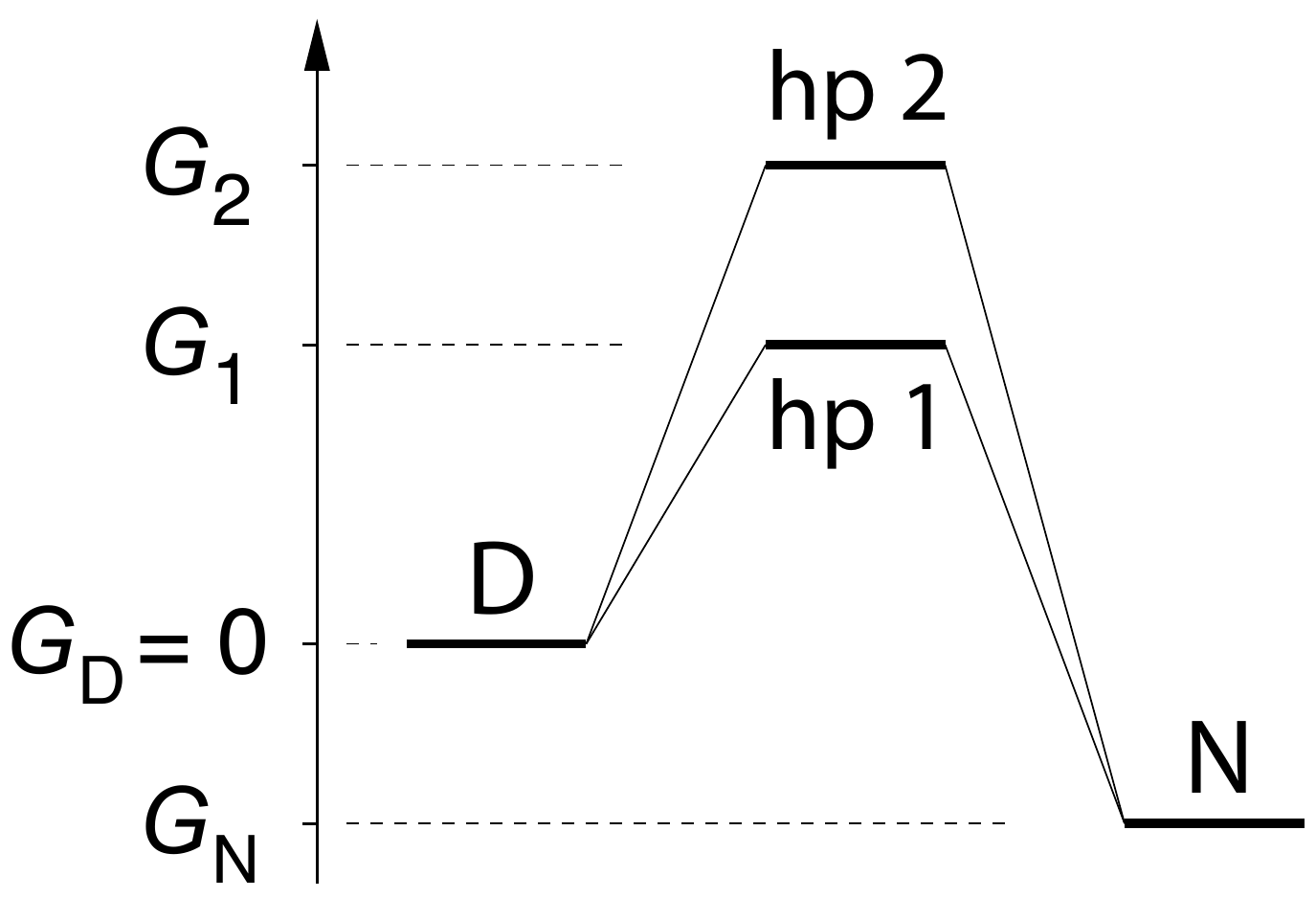}}
\end{center}
\vspace{0.5cm}
\caption{Simple energy landscape of the four-state model for WW domains. The four states are the denatured state $D$, the native state $N$, and two transition-state conformations hp 1 and hp 2 in which one of the two hairpins is formed. Here, $G_N$ is the free-energy difference between the native state $N$ and the denatured state $D$, which has the `reference free energy' $G_D = 0$, and  $G_1$ and $G_2$ are the free energy differences between the transition-state conformations and the denatured state.}
\label{figure_landscape}
\end{figure}

\clearpage

\begin{figure}
\begin{center}
\resizebox{0.55\linewidth}{!}{\includegraphics{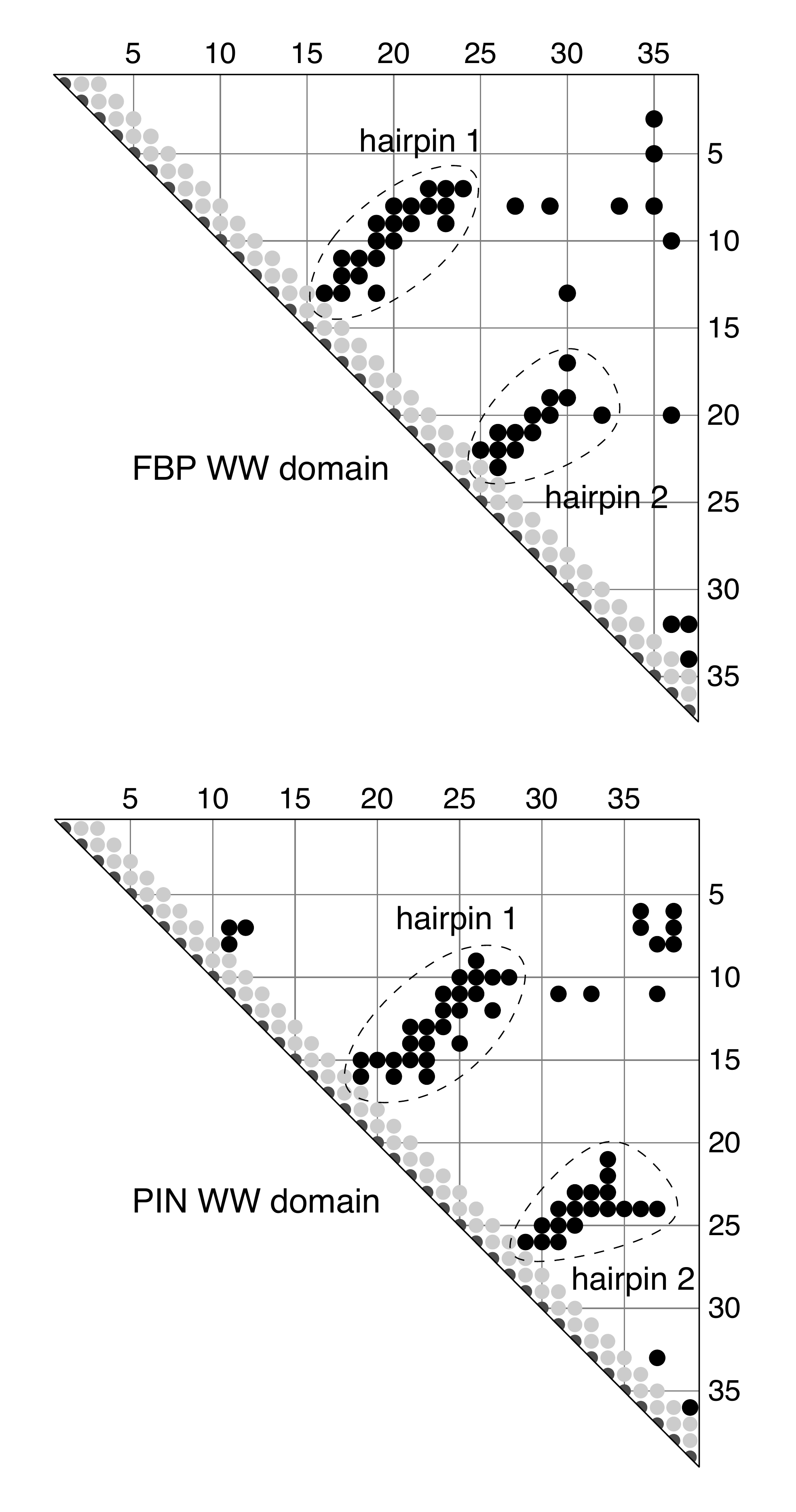}}
\end{center}
\vspace{0.5cm}
\caption{Contact matrices of the FBP and PIN WW domains. A black dot at position $(i,j)$ of a matrix indicates that the residues $i$ and $j$ are in contact. Two residues are defined here to be in contact if the distance between any of their non-hydrogen atoms is smaller than the cutoff distance 4 \AA. The hairpins 1 and 2 of the WW domains correspond to clusters of contacts.}
\label{figure_contactmaps}
\end{figure}

\clearpage

\begin{figure}
\begin{center}
\resizebox{0.9\linewidth}{!}{\includegraphics{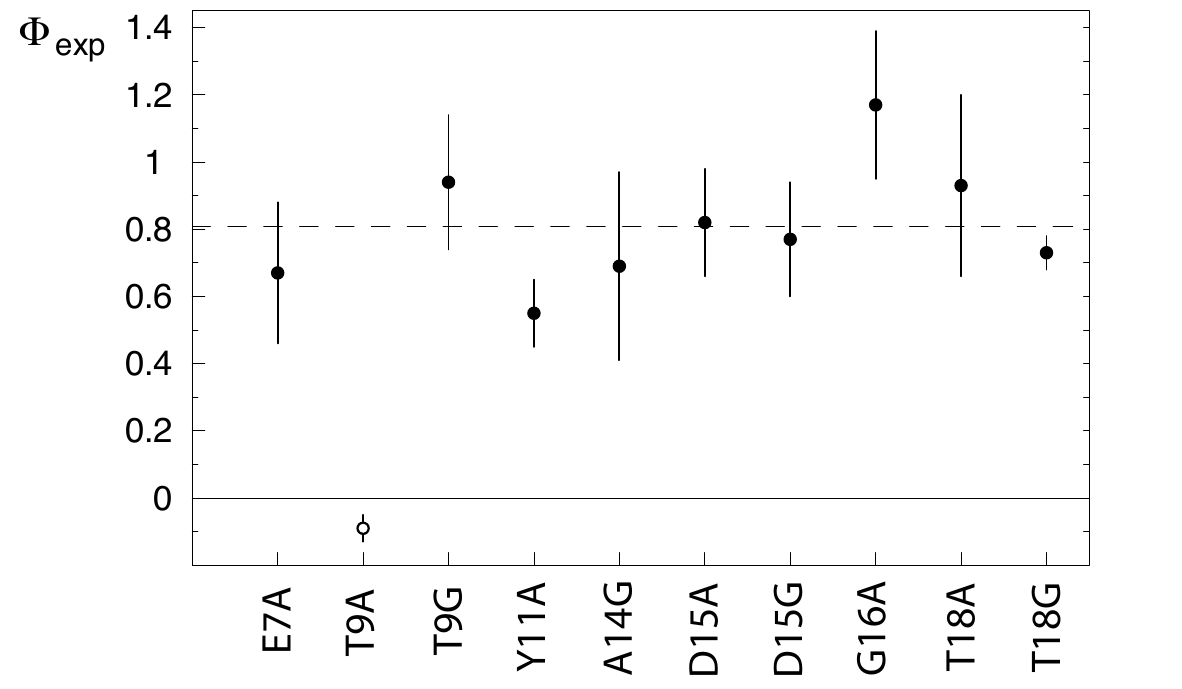}}
\end{center}
\vspace{0.5cm}
\caption{$\Phi$-values for mutations that only affect haipin 1 of the FBP WW domain (see also Table 1). Except for one outlier (open circle for mutation T9A), the $\Phi$-values are centered around the mean value $0.81\pm 0.06$, with deviations mostly within the estimated experimental errors \cite{Petrovich06}. 
}
\label{figure_FBPhairpin1}
\end{figure}

\clearpage

\begin{figure}
\begin{center}
\resizebox{0.9\linewidth}{!}{\includegraphics{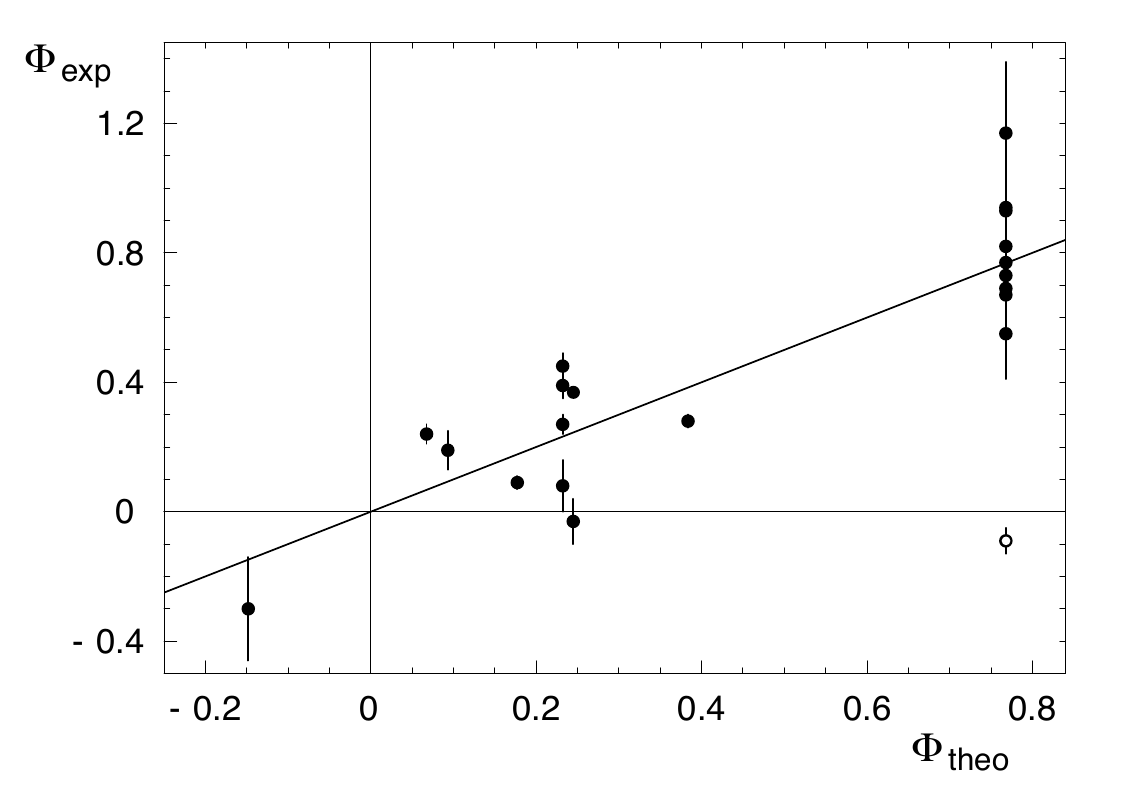}}
\end{center}
\vspace{0.5cm}
\caption{Experimental $\Phi$-values for the FBP WW domain versus theoretical $\Phi$-values obtained from a least-square fit of eq.~(\ref{phi}) with the single fit parameter $\chi_1$. From this fit, we obtain the values $\chi_1=0.77\pm 0.05$ and $\chi_2=1-\chi_1=0.23\pm 0.05$ for the fractions of the two transition-state conformations in which either hairpin 1 or hairpin 2 are formed. The Pearson correlation coefficient between theoretical and experimental $\Phi$-values is $r=0.90$ if the outlier data point for mutation T9A (open circle) is not considered, and $r=0.77$ if the outlier is included.
}
\label{figure_phisfbp}
\end{figure}

\clearpage

\begin{figure}
\begin{center}
\resizebox{0.9\linewidth}{!}{\includegraphics{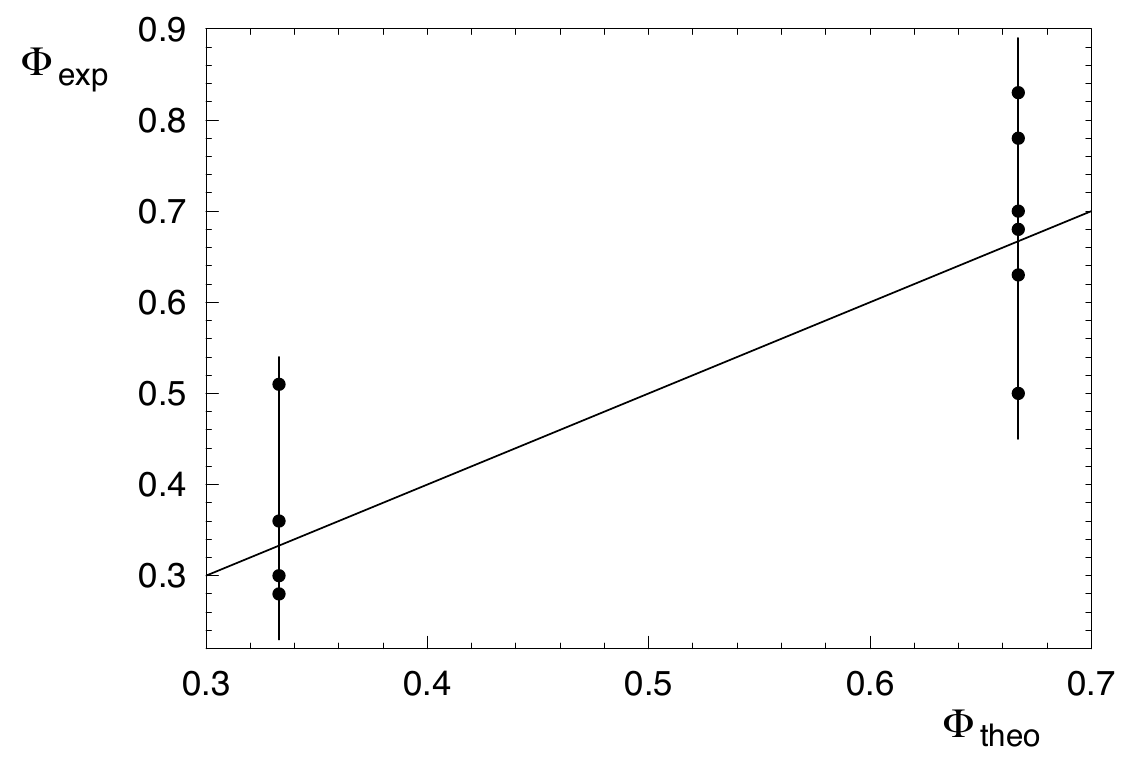}}
\end{center}
\vspace{0.5cm}
\caption{Experimental $\Phi$-values for the PIN WW domain versus theoretical $\Phi$-values obtained from a least-square fit of eq.~(\ref{phi}), which results in the values $\chi_1=0.67\pm 0.05$ and $\chi_2=1-\chi_1=0.33\pm 0.05$ for the fractions of the two transition-state conformations. The Pearson correlation coefficient between theoretical and experimental $\Phi$-values is $r=0.85$.
}
\end{figure}

\end{document}